\documentstyle[twocolumn,psfig]{mn}

\title{On the masses of black-holes in radio-loud quasars}
\author[M.F. Gu, X.W. Cao \& D.R. Jiang]
       {Minfeng Gu, Xinwu Cao and D. R. Jiang \\
1.Shanghai Astronomical Observatory, Chinese Academy of Sciences,
Shanghai, 200030, China; gumf@center.shao.ac.cn\\ 2.National
Astronomical Observatory, Chinese Academy of Sciences, Beijing,
China}
\pagerange{\pageref{firstpage}--\pageref{lastpage}}
\pubyear{}

\begin{document}
\maketitle
\label{firstpage}

\begin{abstract}
The central black-hole masses of a sample of radio-loud quasars
are estimated by using the data of $H_{\beta}$ line-width and the
optical continuum luminosity. The vast majority of the quasars in
this sample have black-hole masses larger than $10^{8} M_{\odot}$,
while a few quasars may contain relatively smaller black-holes. We
found a significant anti-correlation between the radio-loudness
and the central black-hole mass. It might imply that the jet
formation is governed by the black-hole mass.
\end{abstract}

\begin{keywords}
galaxies: active - galaxies: nuclei - quasars: general
\end{keywords}

\section{Introduction}
The radiant energy is believed to be released through accretion
onto supermassive black-holes in the currently most favoured model
of powering quasars. The black-hole mass is a crucial quantity in
this model. There are some different approaches adopted to
estimate the central black-hole mass, such as, gas kinematics
near a black hole, accretion disc spectra fitting, variability in
optical or X-ray wavebands (see Ho \& Kormendy 2000 for a review
and references therein). Due to the technical difficulties on the
studies of the nuclear gas kinematics and the stellar dynamics in
quasars (Nelson 2000), the central black-hole masses are derived
only for a few quasars until recently.

On the assumption of the broad emission lines being produced in
clouds which is gravitationally bound and orbiting with Keplerian
velocities (Dibai 1981), the central black-hole mass can be
derived by using the broad line width and the distance of the
Broad Line Region (BLR) from the centre. The central black-hole
mass is then given by $M_{\rm bh}=R_{\rm BLR}V^{2}G^{-1}$, where
$R_{\rm BLR}$ is the radius of the BLR and $V$ is the velocity of
the clouds in the BLR. This approach is strongly supported by the
Keplerian dynamics of the BLR in a well explored source NGC5548
(Peterson \& Wandel 1999; 2000).

The reverberation-mapping method (Peterson 1993; Netzer \&
Peterson 1997) is applied to measure the radius of the BLR from
the time delay between the line and continuum variations. The
masses of the black-holes in several tens Seyfert galaxies and
quasars have been derived in this way (Wandel et al. 1999; Kaspi
2000). Generally speaking, the distances of the clouds from the
black-hole measured by the reverberation-mapping method are very
reliable, but it can be used only for a few long-term monitoring
objects. Recently, a tight correlation is found between the
radius of the BLR $R_{\rm BLR}$ and the optical continuum
luminosity $L$ (Wandel et al. 1999; Kaspi et al. 1996). Kaspi et
al. (2000) suggested a $R_{\rm BLR}$-$L$ relation: $R_{\rm BLR}
\propto L^{0.7}$, for a sample of Seyfert 1 galaxies and quasars,
in which the radius of the BLR are measured with the
reverberation-mapping method. So, the $R_{\rm BLR}$ can be
derived from this empirical relation, and one can therefore
estimate the central black hole mass of the quasar using observed
broad-line width and optical luminosity.

Franceschini, Vercellone \& Fabian (1998) found a surprisingly
tight correlation between black-hole mass and radio power in a
small sample of nearby mostly non-active galaxies. This relation
is confirmed by Mclure et al.(1999). They measured the properties
of the host galaxies for a sample of Active Galactic Nuclei (AGNs)
including both radio-loud and radio-quiet sources. The black-hole
masses are estimated using the $M_{\rm bh}-M_{\rm bulge}$ relation
suggested by Magorrian et al.(1998). They found that the
black-holes are systematically heavier in radio-loud sources than
that in their counterparts. Mclure \& Dunlop (2000) found the
median black-hole mass of radio-loud quasars is a factor of three
larger than that of radio-quiet quasars in their sample. It is
suggested that a radio-loud quasar may require a black-hole mass
$>6\times10^{8}M_{\odot}$ for some unknown reasons. Recently, Laor
(2000a) found  the similar results for a sample of 87 PG quasars.
He found that most quasars (after excluding four radio
intermediate quasars) with $M_{\rm bh}>10^{9}M_{\odot}$ are
radio-loud and essentially all quasars with $M_{\rm
bh}<3\times10^{8}M_{\odot}$ are radio-quiet.

In this work, we focus on radio-loud quasars. The
central black hole masses are estimated for a sample of
radio-loud quasars. The sample are described in Sect. 2. Section 3
contains the results. The last section is devoted to discussion.
The cosmological parameters $H_{0}=75$ km s$^{-1}$ Mpc$^{-1}$ and
$q_{0}=$0.5 have been adopted in this work.

\section{The  sample}

We consider the sample of a combination of all
quasars in 1 Jy, S4 and S5
catalogues. Optical counterparts have been found for 97 \% of the radio
sources in 1 Jy catalogue (Stickel, Meisenheimer \& K\"uhr 1994).
The S4 survey has a flux density $S_{\rm 5 GHz}\ge 0.5$ Jy
(Pauliny-Toth et al. 1978), while the S5 survey contains the sources
with $S_{\rm 5 GHz}\ge 0.25$ Jy (K\"uhr et al. 1981).
For the S4 catalogue, about 90 \% of the radio sources have known optical
counterparts (Stickel \& K\"uhr 1994), and about 75 \% of the sources
have optical counterparts in S5 catalogue (Stickel \& K\"uhr 1996).
In this combined sample, there are 358 sources identified as quasars.

We search the literature and collect all sources in this sample
with available data of H$_\beta$ profiles. This leads to 86
sources including 55 flat-spectrum sources and 31 steep-spectrum
sources. We find 78 sources with $M_{V}<-23.0$ in this sample.
When more than one value of the H$_{\beta}$ line width are found
in the literature, we take the data in the most recent reference
uniformly. In Table 1., the FWHM (full width at half-maximum) of
H$_{\beta}$ and corresponding references are listed. It appears
that the FWHM of $H_{\beta}$ spreads a wide range from less than
$1000 ~\rm{km~s^{-1}}$ to more than $10000 ~\rm{km~s^{-1}}$.

In this work, we use the relation between $R_{\rm BLR}$ and the
monochromatic luminosity at 5100\AA \hspace{1 mm} found by Kaspi
et al. (2000) to derive the radius of the BLR. We take
$V=1.5\times {\rm H}_{\beta}~{\rm  FWHM}$, suggested by Mclure \&
Dunlop (2000), to derive the velocity of the clouds. The
black-hole mass is then estimated by using $M_{\rm bh}=R_{\rm
BLR}V^{2}G^{-1}$. The results are given in Table 1.

\begin{table}
 \begin{minipage}{120mm}
  \caption{Data of the sample.}
  \begin{tabular}{cccclc}\hline
Source & Rad.Type & z & ${H\beta}_{\rm FWHM}$ &  Refs. &$
Log(M_{\rm bh}/M_{\odot})$
\\ (1) & (2) & (3) & (4) & (5) &   (6) \\ \hline
 0022$-$297& SS &0.406& 4200& B99 & 8.391\\
 0024+348& FS &0.333& 546 &SK93b & 6.852 \\
 0056$-$001& FS &0.717& 3000& B96 & 9.190 \\
 0110+495& FS &0.395& 3628 &H97& 8.821 \\
 0119+041& FS &0.637& 4444&JB91& 8.855\\
 0133+207& SS &0.425& 13500& N95& 10.00 \\
 0133+476& FS &0.859& 2697& L96& 9.206 \\
 0134+329& SS &0.367& 3360& C97& 9.222 \\
 0135$-$247& FS &0.831& 4500& B99& 9.612 \\
 0159$-$117& SS &0.669& 4500& B96& 9.750 \\
 0210+860& SS &0.186& 901& L96& 7.014 \\
 0237$-$233& SS &2.224& 931& B94& 8.998 \\
 0336$-$019& FS &0.852& 4875.5& JB91& 9.459\\
 0403$-$132& FS &0.571& 4780& C97& 9.548 \\
 0405$-$123& FS &0.574& 3590 &C97& 9.946 \\
 0420$-$014& FS &0.915& 3000 & B96& 9.510 \\
 0437+785& SS &0.454& 4486& S93& 9.269 \\
 0444+634& FS &0.781& 3427& SK93b& 9.010 \\
 0454$-$810& FS &0.444 & 4107& S89& 8.608 \\
 0514$-$459& FS &0.194 & 2021& S93& 8.023 \\
 0518+165& SS &0.759& 2840& GW94& 9.004 \\
 0538+498& SS &0.545& 8924& L96& 10.06 \\
 0602$-$319& SS &0.452& 8500& W86 & 9.499 \\
 0607$-$157& FS &0.324& 3517.8& H78& 9.162\\
 0637$-$752& FS &0.654& 4082& AW91& 9.891\\
 0646+600& FS &0.455& 6731& SK93b& 9.220 \\
 0723+679& SS &0.846&  3178& L96& 9.151 \\
 0736+017& FS &0.191&  2820& C97& 8.677 \\
 0738+313& FS &0.631&  4800& B96& 9.880 \\
 0809+483& SS &0.871&  1259& L96& 8.435 \\
 0838+133& SS &0.684&  3090& M96& 8.998 \\
 0906+430& FS &0.668&  1833& L96& 8.378 \\
 0923+392& FS &0.698&  7200 & B96& 9.756 \\
 0953+254& FS &0.712&  4012 &JB91& 9.476 \\
 0954+556& FS &0.901& 1432& L96& 8.549 \\
 1007+417& SS &0.6123&  2860& M96& 9.270 \\
 1034$-$293& FS &0.312&  4116& S89& 9.226 \\
 1045$-$188& FS &0.595&  623& S93& 7.308 \\
 1100+772& SS &0.3115&  6160& C97& 9.783 \\
 1111+408& SS &0.734& 13500&N95& 10.30 \\
 1136$-$135& FS &0.554&  2670& C97& 9.260\\
 1137+660& SS &0.6563&  4950 &M96& 9.840 \\
 1150+497& FS &0.334& 4810 & B96& 9.206 \\
 1151$-$348& SS &0.258&  9430&W86& 9.501 \\
 1202$-$262& FS &0.789&  7500&B99& 9.478 \\
 1226+023& FS &0.158&  3520& C97& 9.693 \\
 1244$-$255& FS &0.633&  4600& B99& 9.517 \\
 1250+568& SS &0.321&  4560& B96& 8.902 \\
 1253$-$055& FS &0.536&  3100& WB86& 8.912 \\
 1302$-$102& FS &0.286&  3400& C97& 9.364 \\
 1354+195& FS &0.720&  4400& B96& 9.920 \\
 1355$-$416& SS &0.313&  10170& C97& 10.21 \\
 1451$-$375& FS &0.314&  4610& C97& 9.297 \\
 1458+718& SS &0.905&  3000& B96& 9.462 \\
 1510$-$089& FS &0.361&  3180& C97& 9.130 \\
 1546+027& FS &0.412&  4700& WB86 & 9.200 \\
 1611+343& FS &1.401& 5600& N95& 10.05 \\
 1634+628& SS &0.988&  1370& L96& 7.762 \\
 1637+574& FS &0.750&  4620&  M96& 9.662 \\
 1641+399& FS &0.594&  5140& C97& 9.901 \\
 1642+690& FS &0.751&  1845& L96& 8.237 \\
 1656+053& FS &0.879&  5000& B96& 10.10 \\
 1704+608& SS &0.371& 6560& C97& 10.04 \\
\end{tabular}
\end{minipage}
\end{table}

\begin{table}
 \begin{minipage}{120mm}
  \contcaption{Data of the sample.}
  \begin{tabular}{cccclc}\hline
  1721+343& FS &0.206&  2880& C97& 8.516 \\
 1725+044& FS &0.293&  2090& C97& 8.545 \\
 1726+455& FS &0.714&  2953& H97& 8.695 \\
 1828+487& SS &0.691& 9159& L96& 10.33 \\
 1849+670& SS &0.657&  6317& SK93a& 9.618 \\
 1856+737& FS &0.460&  4777& H97& 9.366 \\
 1928+738& FS &0.302&  3360& C97& 9.386 \\
 1945+725& SS &0.303&  512& SK93a& 6.957 \\
 2043+749& SS &0.104&  11000& C97& 10.10 \\
 2111+801& FS &0.524&  5443& SK93a& 9.212 \\
 2128$-$123& FS &0.501&  7050& C97& 10.09 \\
 2135$-$147& SS &0.200&  8460& C97& 9.867 \\
 2155$-$152& FS &0.672&  1650& S89& 8.065 \\
 2201+315& FS &0.298& 3380& C97& 9.347 \\
 2216$-$038& FS &0.901&  3300& N95& 9.718 \\
 2218+395& SS &0.655&  694& SK93b& 7.618 \\
 2247+140& FS &0.237&  2520& C97& 8.524 \\
 2251+158& FS &0.859&  2800& N95& 9.644 \\
 2255$-$282& FS &0.926&  3600& B99& 9.641 \\
 2311+469& SS &0.741&  5884& SK93b& 9.780 \\
 2342+821& SS &0.735&  1302& L96& 7.783 \\
 2344+092& FS &0.673&  3900& B96& 9.788 \\
 2345$-$167& FS &0.576&  4999& JB91& 9.198 \\
\hline
\end{tabular}
\end{minipage}

\medskip

Column(1): IAU source name. Column (2): the radio type (FS:
flat-spectrum sources; SS: steep-spectrum sources). Column (3):
redshift. Column (4): H$_{\beta}$ FWHM in unit of km s$^{-1}$.
Column (5): references for H$_{\beta}$ FWHM. Column (6): the
derived black-hole mass.

\vskip 1mm References: AW91: Appenzeller \& Wagner (1991). B94:
Baker et al. (1994). B99: Baker et al. (1999). B96: Brotherton
(1996). C97: Corbin (1997). GW94: Gelderman \& Whittle (1994).
H78: Hunstead et al. (1978). H97: Henstock et al. (1997). JB91:
Jackson \& Browne (1991). L96: Lawrence et al. (1996). M96:
Marziani et al. (1996). N95: Netzer et al. (1995). S89: Stickel
et al. (1989). SK93a: Stickel \& K\"uhr (1993a). SK93b: Stickel
\& K\"uhr (1993b). S93: Stickel et al. (1993). W86: Wilkes
(1986). WB86: Wills \& Browne (1986). ZO90: Zheng \& O'Brien
(1990).

\end{table}

\section{Results}

The radio luminosity is K-corrected to the rest frame of the
sources. The relation between the radio luminosity and the
black-hole mass is present in Fig. 1. It is shown that the vast
majority of quasars have black-hole masses larger than
$10^{8}M_{\odot}$ and some even  $>10^{10}M_{\odot}$. There is a
tendency that the radio luminosity increases with the black-hole
mass, which is consistent with that given by Mclure et al. (1999)
and Laor (2000a).

The radio loudness $R$ and black-hole mass relation is plotted in
Fig. 2. We find a significant anti-correlation between them at
99.98 per cent confidence (Spearman correlation coefficient
$\rho$) with a correlation coefficient: -0.4. It should be
cautious that such a correlation might be caused by the common
dependence of the optical continuum luminosity
$L_{\lambda}$($5100\rm{\AA}$). We therefore use the partial
Spearman rank correlation method (Macklin 1982) to check this
correlation. A relatively weaker correlation with a correlation
coefficient -0.205 is present independent of
$L_{\lambda}$($5100\rm{\AA})$. The significance of the partial
rank correlation is -1.886, which is equivalent to the deviation
from a unit variance normal distribution if there is no
correlation present. This anti-correlation still holds though it
becomes relatively weaker while subtracting the affection of
$L_{\lambda}$($5100\rm{\AA})$.

We use $L_{\rm bol}\approx 9\lambda L_{\lambda}(\rm{5100\AA})$
(Kaspi et al., 2000) to estimate the bolometric luminosity
roughly. Figure 3. shows the distribution of sources in $M_{\rm
bh}$-$L_{\rm bol}$ space. The relation between radio loudness $R$
and $L_{\rm bol}/L_{\rm Edd}$ is plotted in Fig. 4. No significant
correlation is found between them.

\begin{figure}
\centerline{\psfig{figure=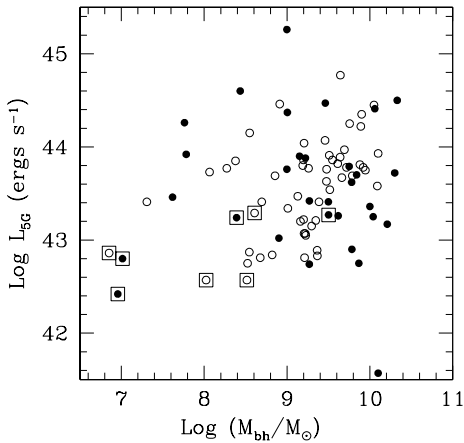,height=6.5cm}} \caption{The
radio luminosity $\nu L_{\nu{\rm 5G}}$ and black-hole mass
relation for the sample. The open circles represent flat-spectrum
sources. The filled circles represent steep-spectrum sources. The
open squares represent the sources with $M_{V}>-23$. }
\end{figure}

\begin{figure}
\centerline{\psfig{figure=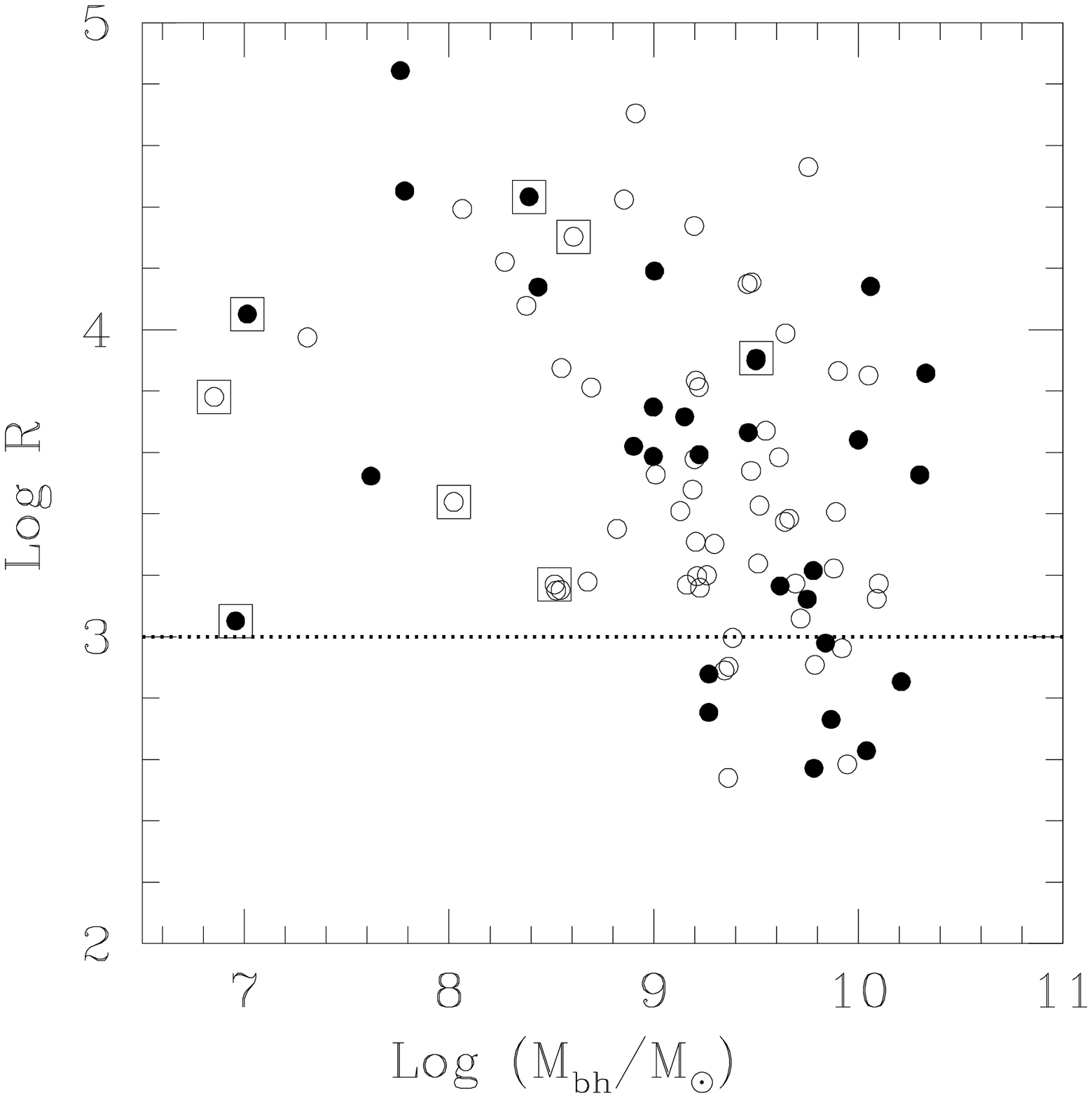,height=6.5cm}} \caption{The
radio loudness parameter $R$ and black-hole mass relation for the
sample. The dotted line represent $R=1000$ (Symbols as in Fig.
1.). }
\end{figure}

\begin{figure}
\centerline{\psfig{figure=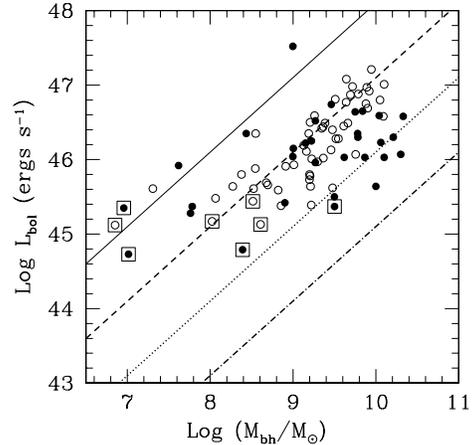,height=6.5cm}} \caption{The
$L_{\rm bol}$ and black-hole mass relation. The four lines
indicate the location of objects radiating with a luminosity equal
to $L_{\rm Edd}$(Solid), $0.1L_{\rm Edd}$(dashed), $0.01L_{\rm
Edd}$(dotted) and $0.001L_{\rm Edd}$(dot-dashed). Symbols are the
same as Fig. 1.}
\end{figure}

\begin{figure}
\centerline{\psfig{figure=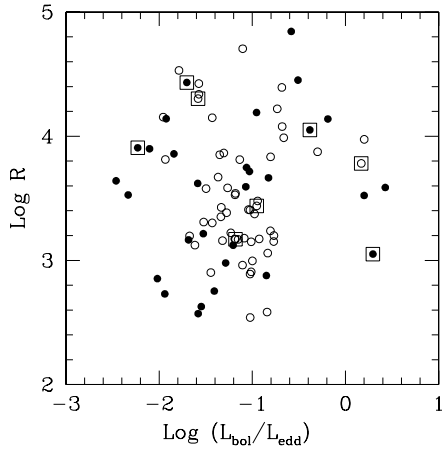,height=6.5cm}} \caption{The
radio loudness $R$ and $L_{\rm bol}/L_{\rm Edd}$ relation
(symbols as in Fig. 1.).}
\end{figure}

\begin{figure}
\centerline{\psfig{figure=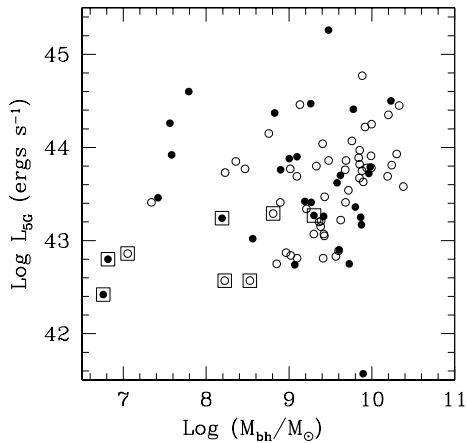,height=6.5cm}} \caption{The
radio luminosity $\nu L_{\nu{\rm 5G}}$ and black-hole mass
relation for the sample. The black-hole masses are re-estimated
after the orientation effect correction of FWHM H$_{\beta}$ by a
factor of ${R_{c}}^{0.1}$ (see text), where $R_{c}$ is the ratio
of core to extended radio fluxes (symbols as in Fig. 1.). }
\end{figure}

\begin{figure}
\centerline{\psfig{figure=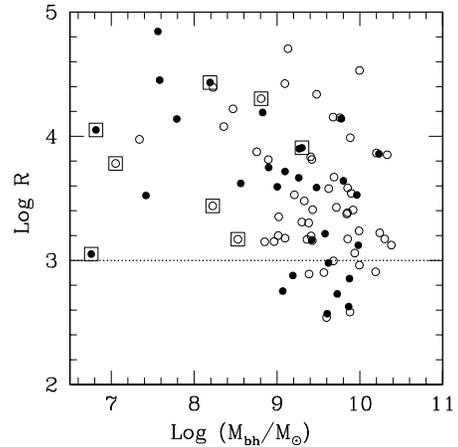,height=6.5cm}} \caption{The
radio loudness parameter $R$ and black-hole mass relation for the
sample. The dotted line represent $R=1000$ (Black-hole mass as in
Fig. 5. and symbols as in Fig. 1.). }
\end{figure}

\section{Discussion}

There are 7 sources in our present sample having black-hole masses
less than  $10^{8}M_{\odot}$. It is unlike the previous
conclusion that all radio-loud quasars (or galaxies) contain
black-holes heavier than $\sim 10^{9}M_{\odot}$ (Mclure et al.
1999; Mclure \& Dunlop 2000; Laor 2000a). It should be noted that
5 of these 7 sources have the H$_\beta$ FWHM less than 1000 km
s$^{-1}$, and the H$_\beta$ FWHM of the left two sources are
slightly larger than 1000 km s$^{-1}$. We note that 3 of these 7
sources are less luminous than $M_{V}=-23.0$. It is obviously that
the narrow $H_\beta$ line leads to a small back-hole for these 7
sources. It is strongly desired to be observed perhaps with $HST$
on these sources to see if the host galaxies of them are
elliptical or not.

The widths of broad H$_\beta$ lines in the sources of our sample
range from a minimum of FWHM $\approx 500~\rm{km~ s^{-1}}$ (only
slightly broader than the narrow lines) to $\geq 10^{4}~\rm{km~
s^{-1}}$. The typical value are around $5000~\rm{km~ s^{-1}}$.
The FWHM of narrow emission lines are usually in the range from
$200~ \rm{km~ s^{-1}}$ to $900~ \rm{km~ s^{-1}}$, and most of
them are around $350-400~ \rm{km~ s^{-1}}$ (Peterson 1997). We
find that the FWHM of H$_\beta$ line of these seven sources
approximate to that of the typical narrow lines (eg. O$[{\sc
iii}]$).
So, we cannot rule out the possibility that the H$_{\beta}$ line
in these sources may be the narrow component emitted from the
narrow line regions. If this is the case, the present procedure of
black-hole mass estimate may be problematic for these sources. So,
the physical origin of these narrow H$_\beta$ lines is crucial in
estimate of black-hole masses for these sources.

Recently, Mclure \& Dunlop (2000) considered a disc-like BLR. The
disc orientation to the observer has been taken into account to
estimate the central black-hole mass. They found that the
estimated virial black-hole masses are quite reliable compared
with that obtained from the $M_{bh}-\sigma$ relation of Gebhardt
et al. (2000) and Merritt \& Ferrarese (2000). For this disc-like
BLR model, the real velocity of the clouds would be larger than
the velocity derived directly from the observed line width if the
disc is face-on. It implies that the masses of the black-hole in
flat-spectrum sources might be under-estimated. The ratio $R_{c}$
of the core to extended radio luminosity in the rest frame of the
sources can be used as an indicator of the jet orientation. We
tentatively correct the orientation effect of the disc-like BLR by
multiplying $H_{\beta}$ FWHM by a factor of ${R_{c}}^{0.1}$ (Lacy
et al. 2001). The value of the $R_{c}$ for most sources in our
sample can be found in Cao \& Jiang (2001). For those sources
without measured $R_{c}$, we assume $R_{c}=0.1$ for
steep-spectrum sources and $R_{c}=10$ for flat-spectrum sources,
respectively (Lacy et al. 2001). We can then re-estimate the
black-hole masses for the sources in our sample. The black-hole
masses are ploted against the radio luminosity and the
radio-loudness in Fig. 5 and Fig. 6, respectively.

Both Figs. 1 and 5 show a tendency that the radio luminosity
increases with the black-hole mass. It still exists if we exclude
seven sources with black-hole masses $M_{bh}<10^{8}M_{\odot}$. It
is consistent with the results given by Mclure et al.(1999) and
Laor(2000a). We find that the correlation has been improved in
Fig. 5., where the orientation effect on the velocity of clouds
is tentatively corrected. In some theoretical jet formation
models, the power is generated through accretion and then
extracted from the disc/black-hole rotational energy and
converted into the kinetic power of the jet (Blandford \& Znajek
1977; Blandford \& Payne 1982). Roughly speaking, the jet
formation is predominantly set by the black-hole (Laor 2000b),
though the details are still unknown. It is easy to understand
that the more massive black-holes always channel much of the
accretion gas into powerful relativistic jets, then the larger
radio luminosity appears.

The radio loudness parameter $R\equiv
f_{\nu}$(5GHz)$/f_{\nu}(\rm{4400\AA})$ is  a good indicator of
the ratio of jet power to accretion power, at least for
steep-spectrum quasars. The anti-correlation between radio
loudness $R$ and the black-hole mass implies that the jet
formation is related with the black-hole mass. For flat-spectrum
quasars, the radio emission is strongly beamed to us, and the
optical emission may also be contaminated by the synchrotron
emission from the jet. We then limit our analysis on
steep-spectrum quasars in our sample and find a similar
anti-correlation as the whole sample. The bolometric luminosity
is estimated from the optical luminosity in this work. So, the
masses of black-holes in flat-spectrum quasars may be
over-estimated, i.e., masses given in Table 1 are the upper limits
for flat-spectrum quasars if the optical continuum emission
contains emission from the jets. We note that all sources with
radio loudness $R<1000$ have black-hole masses larger than
$10^9M_{\odot}$, which is consistent with Laor's (2000a) result.

In Fig. 3., we find that the vast majority of the objects appear
to be radiating between 0.01 and 0.1 Eddington luminosity. We
note that there is no correlation between radio loudness $R$ and
luminosity $L_{\rm bol}/L_{\rm Edd}$. It may implies that the jet
formation is not related to the type of accretion disc in the
source.

\section*{Acknowledgments}
We thank the anonymous referee for the helpful comments. The
support from NSFC, the NKBRSF (No. G1999075403), and Pandeng
Project is gratefully acknowledged. This research has made use of
the NASA/IPAC Extragalactic Database (NED), which is operated by
the Jet Propulsion Laboratory, California Institute of Technology,
under contract with the National Aeronautic and Space
Administration.

{}


\begin{thebibliography}{}
\bibitem{}Appenzeller I., Wagner S.J., 1991, A\&A, 250, 57
(AW91)
\bibitem{}Baker A.C., Carswell R.F., Bailey J.A., Espey B.R., Smith M.G.,
    Ward M.J., 1994, MNRAS 270, 575(B94)
\bibitem{}Baker J.C., Hunstead R.W., Kapahi V.K., Subrahmanya
C.R., 1999, ApJS, 122, 29 (B99)
\bibitem{}Blandford R. D., Payne D. G., 1982, MNRAS, 199, 883
\bibitem{}Blandford R. D., Znajek R. L., 1977, MNRAS, 179, 433
\bibitem{}Brotherton M.S., 1996, ApJS, 102, 1 (B96)
\bibitem{}Cao, X., Jiang D.R., 2001, MNRAS, 320, 347
\bibitem{}Corbin M.R., 1997, ApJS, 113, 245 (C97)
\bibitem{}Dibai E.A., 1981, Soviet Astron., 24, 389
\bibitem{}Franceschini A., Vercellone S., Fabian A.C., 1998, MNRAS, 297, 817
\bibitem{}Gebhardt K. et al., 2000, ApJ, in press (astro-ph/0006289)
\bibitem{}Gelderman R., Whittle M., 1994, ApJS, 91, 491(GW94)
\bibitem{}Henstock D.R., Browne I.W.A., Wilkinson P.N., McMahon
R.G., 1997, MNRAS, 290, 380 (H97)
\bibitem{}Ho L.C., Kormendy J., 2000, The Encyclopedia of Astronomy and Astrophysics (Institute of Physics
Publishing). (astro-ph/0003267)
\bibitem{}Hunstead R.W., Murdoch H.S., Shobbrook R.R., 1978,
MNRAS, 185, 149 (H78)
\bibitem{}Jackson N., Browne I.W.A., 1991, MNRAS, 250, 422 (JB91)
\bibitem{}Kaspi S., Smith P.S., Maoz D., Netzer H., Jannuzi B.T., 1996, ApJ, 471,
L75
\bibitem{}Kaspi S., Smith P.S., Netzer H., Maoz D., Jannuzi B.T.,
Giveon U., 2000, ApJ, 533, 631
\bibitem{}K\"uhr M., Pauliny-Toth I.I.K., Witzel A., Schmidt J., 1981, ApJ, 86, 854
\bibitem{}Lacy M., Laurent-Muehleisen S.A., Ridgway S.E., Becker
R.H., White R.L., (astro-ph/0103087)
\bibitem{}Laor A., 2000a, ApJ, 543, L111
\bibitem{}Laor A., 2000b, (astro-ph/0005144)
\bibitem{}Lawrence C.R., Zucker J.R., Readhead A.C.S., Unwin S.C.,
   Pearson T.J., Xu W., 1996, ApJS, 107, 541 (L96)
\bibitem{}Macklin J.T., 1982, MNRAS, 199, 1119
\bibitem{}Magorrian J., et al. 1998, AJ, 115, 2285
\bibitem{}Marziani P., Sulentic J.W., Dultzin-Hacyan D., Calvani M., Mole
   M., 1996, ApJS, 104, 37(M96)
\bibitem{}Mclure R.J., Dunlop J.S., 2000, MNRAS, in press
(astro-ph/0009406)
\bibitem{}Mclure R.J., Kukula M.J., Dunlop J.S., Baum S.A., O'Dea
C.P., Hughes D.H., 1999, MNRAS, 308, 377
\bibitem{}Merritt D., Ferrarese L., 2000, ApJ, in press
(astro-ph/0008310)
\bibitem{}Nelson C.H., 2000, ApJ, in press (astro-ph/0009188)
\bibitem{}Netzer H., Brotherton M.S., Wills B.J., Han M.Sh., Wills
D., Baldwin J.A., Ferland G.J., Browne I.W.A., 1995, ApJ, 448, 27
(N95)
\bibitem{}Netzer H., Peterson B.M., 1997, in Astronomical Time
Series, ed. Maoz D., Sternberg A. \& Leibowitz E. (Dordrecht:
Kluwer), 85
\bibitem{}Pauliny-Toth I.I.K., Witzel A., Preuss E. et al., 1978, ApJ, 83, 451
\bibitem{}Peterson B.M., 1993, PASP, 105, 207
\bibitem{}Peterson B.M., 1997, An introduction to active galactic
nuclei (Cambridge: Cambridge Univ. Press)
\bibitem{}Peterson B.M., Wandel A., 1999, ApJ, 521, L95
\bibitem{}Peterson B.M., Wandel A., 2000, ApJ, 540, L13
\bibitem{}Stickel M., Fried W., K\"uhr H., 1989, A\&AS, 80, 103(S89)
\bibitem{}Stickel M., K\"uhr H., 1993a, A\&AS, 100, 395 (SK93a)
\bibitem{}Stickel M., K\"uhr H., 1993b, A\&AS, 101, 521 (SK93b)
\bibitem{}Stickel M., K\"uhr H., Fried W., 1993, A\&AS, 97, 483(S93)
\bibitem{}Stickel M., K\"uhr H., 1994, A\&AS, 103, 349
\bibitem{}Stickel M., K\"uhr H., 1996, A\&AS, 115, 1
\bibitem{}Stickel M., Meisenheimer K., K\"uhr H., 1994, A\&AS, 105, 211
\bibitem{}Wandel A., Peterson B.M., Malkan M.A., 1999, ApJ, 526,
579
\bibitem{}Wilkes B.J., 1986, MNRAS, 218, 331 (W86)
\bibitem{}Wills B.J., Browne I.W.A., 1986, ApJ, 302, 56 (WB86)
\bibitem{}Zheng W., O'Brien P.T., 1990, ApJ, 353, 433, (ZO90)

\end{thebibliography}
\end{document}